# METRICS FOR AERIAL, URBAN LIDAR POINT CLOUDS
*Michael H. Stanley[a] and Debra F. Laefer[b,*]*

[a]Center for Data Science, New York University,
60 5th Avenue, New York, NY 11201, USA
e-mail: michael.stanley@nyu.edu
[b]Center for Urban Science and Progress; Department of Civil Engineering,
Tandon School of Engineering, New York University,
370 Jay Street, 1301C, Brooklyn, NY 11201, USA
email: debra.laefer@nyu.edu
*Corresponding author

ABSTRACT
This paper introduces five new density and accuracy metrics for aerial point clouds that address the complexity and objectives of modern, dense laser scans of urban scenes. The five metrics describe (1) vertical surface density (points per area on vertical surfaces); (2) vertical density as a function of horizontal density; (3) vertical surface accuracy; and a decomposition of error into (4) within-pass and (5) cross-pass components. Specifically considered is vertical surface coverage and the practice of overlapping flight passes to reduce the occlusions and achieve the vertical density needed for twenty-first-century use cases (e.g. curb and window detection). The application of these metrics to a quartet of recent urban flyovers demonstrates their relevance by establishing (1) the efficacy of considering sensor position and wall height when predicting point density on vertical surfaces; (2) that cross-pass registration accounts for a disproportionate amount of the vertical surface error (but not horizontal) and provides a meaningful parameter to compare high-density, urban point clouds; and (3) that compared to horizontal density and accuracy, the vertical counterparts are disproportionately impacted (positively for density and negatively for accuracy) by modern, optimized flight missions.

## 1. INTRODUCTION

While aerial light detection and ranging (LiDAR) [also known as aerial laser scanning] has been commercially available since the 1960s (Petrie and Toth, 2018), its adoption, project scale, and range of applications have expanded rapidly in the past two decades. This is most easily seen in national aerial LiDAR scan proliferation, which has been driven by radical improvements in LiDAR output density. For example, in 2003, The Netherlands undertook its first national scan at 0.1-2 $pts/m^2$, followed by 3 successive surveys at increasingly higher point densities (6-10 $pts/m^2$ in 2012 and in 2019, with 10-14 $pts/m^2$ – planned for completion in 2022). At least 9 other countries have completed national surveys with point densities between 0.5-20 $pts/m^2$ (Table 1). In 2016, the United States Geological Survey (USGS) launched the 3D Elevation Program (3DEP), with the goal of acquiring the first national LiDAR survey in the United States (US) by 2023. Motivation for that project was in part based on a predicted fivefold ($13

billion) return on investment (USGS 2020). In the USGS's most recent annual report, 67% of the US was surveyed or in the process of being surveyed, with a minimum Aggregate Nominal Point Density of 2 $pts/m^2$ (USGS 2020).

| Location | Type | Year Completed | Point Density ($pts/m^2$) | Spatial extent ($km^2$) | Flight AGL (m) | Source |
|---|---|---|---|---|---|---|
| Denmark | National | 2015 | 8 | 43,000 | - | Flatman et al., 2016 |
| Estonia | National | 2011 | 2-3 | 45,000 | 1300-2400 | Estonia Land Board, 2019 |
| Finland | National | 2010 | 0.5 | 338,000 | - | NLS Finland, 2020 |
| Netherlands (AHN-1) | National | 2003 | 0.1-2 | 42,000 | - | AHN, 2020 |
| Netherlands (AHN-2) | National | 2012 | 6-10 | 42,000 | - | AHN, 2020 |
| Netherlands (AHN-3) | National | 2019 | 6-10 | 42,000 | - | AHN, 2020 |
| Netherlands (AHN-4) | National | 2022* | 10-14 | 42,000 | - | AHN, 2020 |
| Poland | National | 2015 | 4-12 | 290,000 | - | GUGIK, 2020 |
| Spain | National | 2015 | 0.5 | 506,000 | - | PNOA, 2020 |
| Slovenia | National | 2015 | 5 | 20,000 | - | ARSO, 2015 |
| Switzerland | National | 2023* | 15-20 | 40,000 | - | Swisstopo, 2020 |
| Sweden - Laserdata NH | National | 2009 | 0.5-1.0 | 450,000 | 1700-2300 | Lantmateriet, 2020 |
| Sweden - Laserdata Skog (forest) | National | 2018 | 1-2 | 337,500 | 3000 | Lantmateriet, 2020 |
| United States | National | 2023* | >2 | 9,834,000 | - | USGS, 2020 |
| Belgium - Flanders DHMV-I | Regional | 2004 | 0.05 | 14,000 | - | Flanders Information Agency, 2006 |
| Belgium - Flanders DHMV-II | Regional | 2015 | 16 | 14,000 | - | Flanders Information Agency, 2015 |
| Vienna, Austria | Municipal | 2007 | 50 | 400 | - | Vo et al., 2016 |
| Duursche, Netherlands | Municipal | 2007 | 70 | 1 | - | Vo et al., 2016 |
| Zeebrugges, Netherlands | Municipal | 2011 | 65 | 1 | 300 | Vo et al., 2016 |
| Dublin, Ireland (2007) | Municipal | 2007 | 225 | 1 | 400 | Laefer et al., 2014 |
| Dublin, Ireland (2015) | Municipal | 2015 | 348 | <2 | 300 | Laefer et al., 2017 |
| Brooklyn, NY, USA | Municipal | 2019 | 570 | 1 | 300-400 | Laefer and Vo, 2020 |

Table 1. Notable Aerial Scans (* indicates projected completion year)

Much denser LiDAR scans in the range of 50-70 $pts/m^2$ have been commissioned at the municipal level (Table 1). These include Vienna, Austria and portions of the cities of Duursche and Zeebrugges in The Netherlands (Vo et al. 2016). Those densities were typically achieved by helicopter at lower flight altitudes [300-400m above ground level ("AGL")] with a single flight pass or minimally overlapping flight passes, as opposed to the fixed-wing aircraft used in the national scans. The stated use cases for denser municipal scans vary but often include disaster recovery and flood risk assessment (NYC DOITT 2018), urban planning and asset management (Höfle and Hollaus 2010), and building modeling (Forlani et al. 2006).

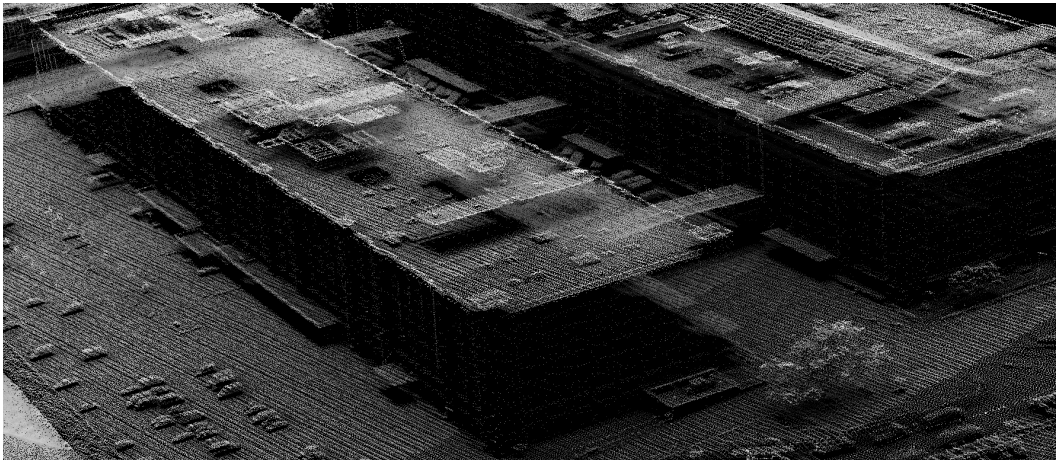
a) Brooklyn Army Terminal from 2017 scan (NYC DOITT 2018)

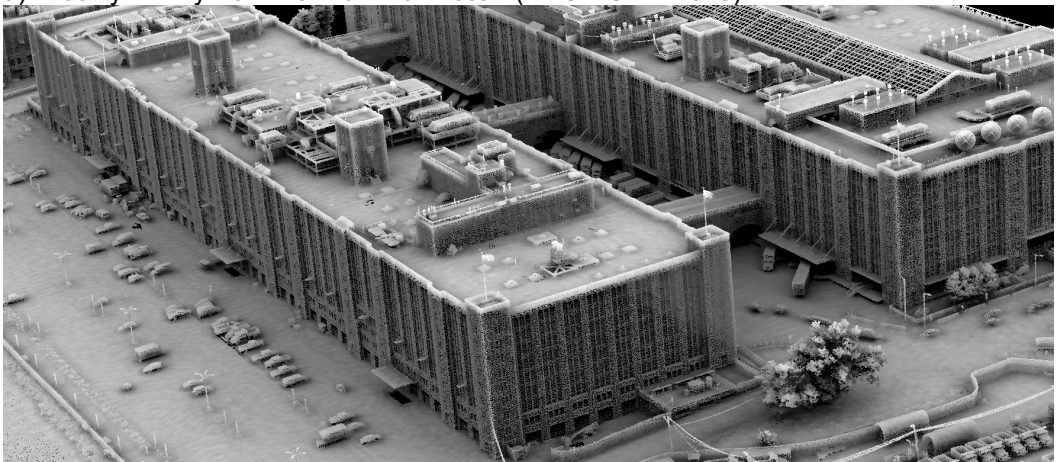
b) Brooklyn Army Terminal from 2019 scan (Laefer and Vo 2020)

Figure 1. Comparison of Aerial LiDAR scans of the Brooklyn Army Terminal. Dark areas are locations without data. Images rendered using ambient occlusion .

Despite increased point densities, the vertical data capture for many municipal efforts remains too sparse for a range of applications such as street curb and utility pole detection (Laefer 2019) or vertical feature classification (e.g. doors and window) for machine learning models (Zolanvari et al. 2019). To that end, a handful of district-scale, urban aerial LiDAR scans have achieved substantially higher point densities, more comprehensive coverage, and improved vertical data density by combining multiple, overlapping flight paths oriented diagonally to the street grid and flown at slower speeds (e.g. 90km/h) [Hinks et al. 2009] and lower heights (e.g. 300-400m AGL) [see Table 1]. One of the first such scans was conducted in 2007 over a 1 $km^2$ area of Dublin, Ireland and combined 44 flight paths for an aggregate point density of 225 $pts/m^2$ (Vo et al. 2016). In 2015, the same team rescanned a slightly expanded area using the same flight specifications but with improved equipment. That scan reached an aggregate point density of 348 $pts/m^2$ over 1.5 $km^2$ (Laefer et al. 2017). Most recently, a 2019 multi-pass scan of a 1 $km^2$ area of Brooklyn, New York achieved an average aggregate point density of 570 $pts/m^2$ (Laefer and Vo 2020) using similar specifications to the Dublin, Ireland scans. Figure 1 shows a traditional, low-density scan (upper) and a recent, high density scan of the same building (lower).

To date, existing density and accuracy metrics for single and multi-pass aerial LiDAR have only considered point density on horizontal surfaces and accuracy of the point cloud as a whole. Such metrics fail to directly address two emerging hallmarks of contemporary urban scanning: vertical surface capture characteristics and the impact of combining numerous flight paths. Arguably, new metrics are needed that explicitly consider vertical surface point density and the error introduced from combining overlapping flight passes (not just strip adjustments). To address these gaps, this paper introduces the following metrics: (1) vertical surface density; (2) vertical density as a function of horizontal density; (3) vertical surface accuracy; and a decomposition of error into (4) within-pass and (5) cross-pass components. These metrics are intrinsic and do not require external ground truth measurements. As part of this metrics development, the angle of capture analysis of Hinks et al. (2009) is extended to include vertical density at various wall heights for a popular commercial scanner type. This paper aims to demonstrate the importance of these metrics by comparing equivalent portions for three contemporary aerial scans.

As the topic of aerial point cloud data acquisition is a large one, several related topics will be considered as outside the scope of this study. These include completeness, absolute accuracy with respect to ground control points, error attribution, and impact of flight path planning on point cloud accuracy.

This paper proceeds with a historical background of aerial LiDAR metrics, practices, and concerns (Section 2), followed by a description of the proposed metrics, the methodology utilized to compare the point clouds, and the datasets (Section 3). The results follow (Section 4), as well as a discussion of the sensitivity and robustness of the new metrics (Section 5). The paper closes with a discussion of potential use cases for high-density LiDAR scanning and future work (Section 6).

2. BACKGROUND
This section provides context around the traditional metrics for point cloud characterization, the methods for performing aerial LiDAR scans, and the impact of those methods on point cloud density and accuracy.

*2.1. Aerial LiDAR scan metrics*
Traditional aerial LiDAR scan metrics have focused on density and accuracy. For multi-pass scan missions, the most common density metrics relate to aggregate counts. These include (1) the aggregate nominal pulse density – the average number of pulses per area of a relatively flat, horizontal surface within the surveyed swath, and (2) the aggregate nominal pulse spacing – the square root of the inverse of the pulse density. Spacing is more common for low-density scans, whereas density is more typical for scans with densities exceeding 1 $pts/m^2$ (Heidemann 2018). Both metrics are dependent on sample surface selection: if the sampled surface is not representative of the dataset overall, then the corresponding metrics will not be representative either.

Density metrics must differentiate between pulses and points. In reporting metrics, the term pulse refers to the first return obtained for each emitted pulse. A single emitted pulse from a LiDAR unit may generate multiple returning points, if the emitted encounters multiple objects on its path groundward and a portion of the pulse reflects off of each object. Because of this phenomenon, the point count is nearly always higher than the pulse count for a given dataset. For example, the 2019 Sunset Park scan listed in Table 1 has 1.06 billion points obtained from 924 million pulses. As most applications use all available points, point-based metrics are usually more useful when profiling existing datasets, while pulse-based metrics are usually used during planning. The remainder of this paper considers aggregate nominal point density (ANPD) as the preferred reporting metric.

The second common metric for aerial LiDAR is accuracy. Quantifying the accuracy of an aerial LiDAR scan is inherently problematic due to the absence of ground truth for the entirety of the scan area. Thus, aerial LiDAR scans typically rely on ground control points, but these are expensive to implement and require access to the ground area being mapped (Habib 2018). Common practice estimates local accuracy, typically described using root mean squared error (RMSE). This approach involves identifying a relatively flat (or at least planar) area and evaluating the point cloud's vertical deviation from that surface (ASPRS 2004). RMSE can only capture error in the vertical direction.

Other evaluation methods focus on quantifying relative accuracy by considering areas of overlap between different point clouds (Latypov 2002). Such a methodology considers a flat surface, G, and the points that lie within the area $A_G$ comprising that surface. The average height for the points of each point cloud is calculated, and the height difference between point clouds is an indicator of closeness. Height differences can be calculated for multiple surfaces, and the statistics (e.g., mean, standard deviation) of those differences provide a measure of relative accuracy between point clouds. This paper does not address error attribution, (e.g., ranging error, angular error, internal measurement unit error) as the subject is well-studied elsewhere (e.g. Glennie 2007).

*2.2 Aerial LiDAR scanning practice*
Historically, aerial LiDAR scanning has employed flight planning and equipment parameters reflecting rural applications. These include data collection from fixed-wing aircraft flown at altitudes of 1km-3km AGL and at relatively high speeds (150-300 km/h) with minimal coverage overlap between adjacent flight strips. These choices tend to generate low point densities (below $10\ pts/m^2$), despite the ability of modern equipment to capture $35 - 50\ pts/m^2$ in a single pass (Riegl 2012).

Point clouds resulting from these data acquisition strategies are sufficiently dense and accurate for large-scale elevation mapping and forest density estimation but do not support many urban applications (e.g. Zolanvari et al. 2019; Vo et al. 2019). For example, aerial LiDAR scans with point densities below $10\ pts/m^2$ are not viable for applications involving sub-building scale objects (e.g. chimneys, steps). When tasked with such applications, local communities have had to rely on terrestrial or mobile LiDAR or ground-based imagery for dense, local point cloud generation either as the sole data set or as a supplementary data set. As an alternative, Hinks et al. (2009) proposed modifications in aerial LiDAR flight paths, to increase aggregate point densities and minimize vertical surface occlusions. Specifically, increasing strip overlap to 67%+, flying at lower speeds and altitudes AGL, and combining multiple flight passes has generated point densities over 500 $pts/m^2$ (Laefer and Vo 2020). By combining the overlapping flight passes collected in this manner, complete surface coverage (which occurs when beam footprints sufficiently overlap to cover the entire surface) can be attained without expanding beam footprints and sacrificing resolution, as described in (Mandlburger et al. 2015). Executing flight paths that are low and diagonal to the grid can be difficult when extremely tall buildings are present sporadically or no grid exists, but this paper does not address these more site-specific aspects of flight planning and, instead, addresses the general and more commonly encountered scenario.

*2.3 Point density as a function of flight parameters*
Flight speed, altitude AGL, and angle of capture all influence aerial LiDAR point density. To most clearly explain their impacts, the following description assumes an even, horizontal surface and a single return per emitted pulse.

LiDAR scanners with a parallel scan pattern emit pulses across their operating range (up to 30° from nadir, for the scanners used to capture the datasets considered herein) at a constant pulse repetition rate, $f_{scanner}$ ($\frac{pts}{s}$). If $A_{scanned}$ ($m^2$) denotes the rectangular area of ground scanned in one second, then point density, $\rho$ ($\frac{pts}{m^2}$), can be calculated:

$$\rho = \frac{f_{scanner}}{A_{scanned}} \tag{eq 1}$$

where $A_{scanned}$ ($m^2$) is the product of two orthogonal components: $d_{across}(m)$ (the end-to-end distance scanned orthogonal to the path of the aircraft) and $d_{along}(m)$ (the distance scanned per second along the aircraft path); $d_{across}$ and $d_{along}$ are wholly determined by flight parameters:

$$d_{across} = 2H \tan(\max(\theta_H))$$
$$d_{along} = v_{aircraft} \tag{eq 2}$$

where H is the aircraft's altitude AGL in meters, $\theta_H$ is the scan angle is in degrees relative to nadir, and $v_{aircraft}$ ($\frac{m}{s}$) is the velocity of the aircraft. Horizontal point density can be calculated as a function of the aircraft's altitude AGL and speed :

$$\rho = \frac{f_{scanner}}{A_{scanned}} = \frac{f_{scanner}}{2 v_{aircraft} H \tan(\max(\theta_H))} \tag{eq 3}$$

Figure 2 illustrates these factors.

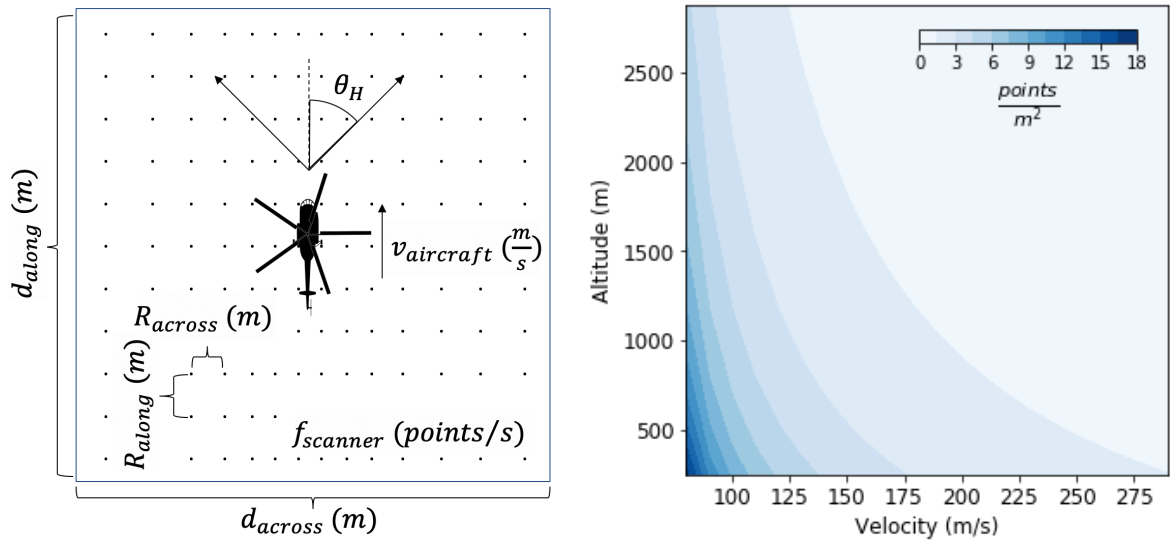

a) Schematic measurements of point density for aerial LiDAR considering points nadir to the aircraft.

b) Point density as a function of flight velocity and scanner AGL based on Riegl Q680i specifications of $f_{scanner} = 400 kHz$, $\theta_H = 30°$.

Figure 2. Impact of flight parameters on point density.

The scan angle (the angle between the emitted pulse and nadir) also impacts point density. Hinks et al. (2009) considered its impact on linear resolution (or point spacing), defined as the distance between consecutive points. This is inversely related to the horizontal point density based on Eq (1):

$$\rho = \frac{f_{scanner}}{A_{scanned}} = \frac{1}{R_{across} R_{along}} \qquad \text{(eq 4)}$$

where $R_{across}(m)$ is the across-path point spacing, and $R_{along}(m)$ is the along-path point spacing. Modern LiDAR scanners can be configured so that $R_{across}$ and $R_{along}$ are roughly equal. Importantly, $R_{across}$ is independent of $R_{along}$ (which is assumed constant for this analysis) which means that $\rho$ is inversely proportional to $R_{across}$.

Figure 3 illustrates the impact of angle of capture on $R_{across}$ at nadir ($R_N$), at angle of capture $\theta_H$ ($R_H$), and at the base of a vertical wall ($R_V$), which can be quantified:

$$R_N = H \tan(\theta_L) \qquad \text{(eq 5)}$$
$$R_H = R_N \sec^2(\theta_H) \qquad \text{(eq 6)}$$
$$R_V = R_H \cot(\theta_H) \qquad \text{(eq 7)}$$

Where H is the aircraft altitude AGL, and $\theta_L$ is the angular step width of the LiDAR scanner (the angle the scanner sweeps between consecutive pulses). Modern scanners typically have $\theta_L < 0.1°$, so the approximations $\sin(\theta_L) \approx 0$ and $\cos(\theta_L) \approx 1$ are used in Eqs (6-12). As the angle of capture, $\theta_H$, increases from 0° at nadir, $R_H$ worsens. If the beam encounters a wall, the vertical resolution at the base of the wall, $R_V$, is worse still.

Critically, if the horizontal offset from the flight path to a wall is small and the aircraft altitude AGL is large, then $\theta_H$ will be small, and $R_V$ will be large. In fact, $R_V$ is infinite at nadir, as the scanner cannot capture a vertical surface positioned directly below the scanner.

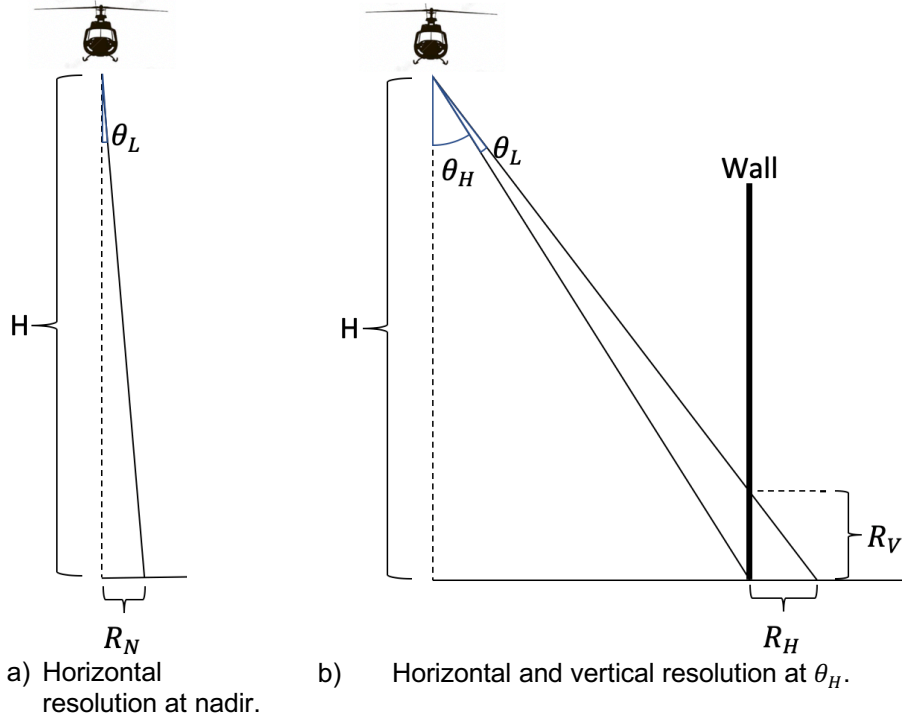

a) Horizontal resolution at nadir.

b) Horizontal and vertical resolution at $\theta_H$.

Figure 3. Impact of angle of capture on point resolution.

Equations (5-7) can be written in terms of point density:

$$\rho_N = \frac{1}{R_{across} R_{along}} = \frac{1}{H \tan(\theta_L) R_{along}} \quad \text{(eq 8)}$$

$$\rho_H = \rho_N \cos^2(\theta_H) \quad \text{(eq 9)}$$

$$\rho_V = \rho_H \tan(\theta_H) \quad \text{(eq 10)}$$

Horizontal density decreases as $\theta_H$ increases, but vertical density increases in the range $0° \leq \theta_H \leq 45°$ before decreasing from $45° \leq \theta_H \leq 90°$ (figure 4). Nadir scanners typically operate in the range $0° \leq \theta_H \leq 30°$, but oblique scanners, can operate across larger scan ranges.

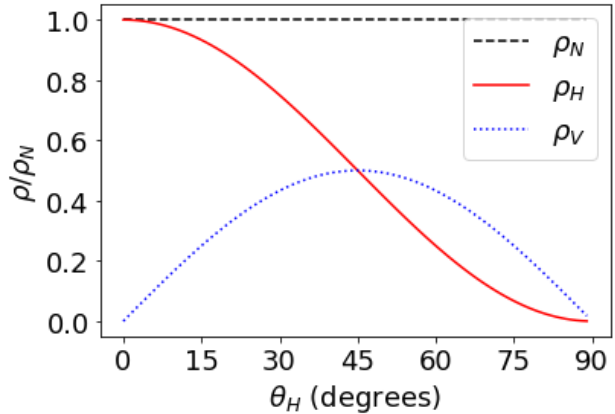

Figure 4. Point densities relative to $\rho_N$ ($0° \leq \theta_H + \theta_L \leq 90°$).

## 3. METHODOLOGY

This section introduces metrics related to general data yield and localized mission accuracy. As part of this, the angle of capture analysis of Hinks et al. (2009) is extended to include vertical density at various wall heights. The workflow used to apply these metrics is then explained, followed by a detailed description of the datasets used to demonstrate the value of the metrics.

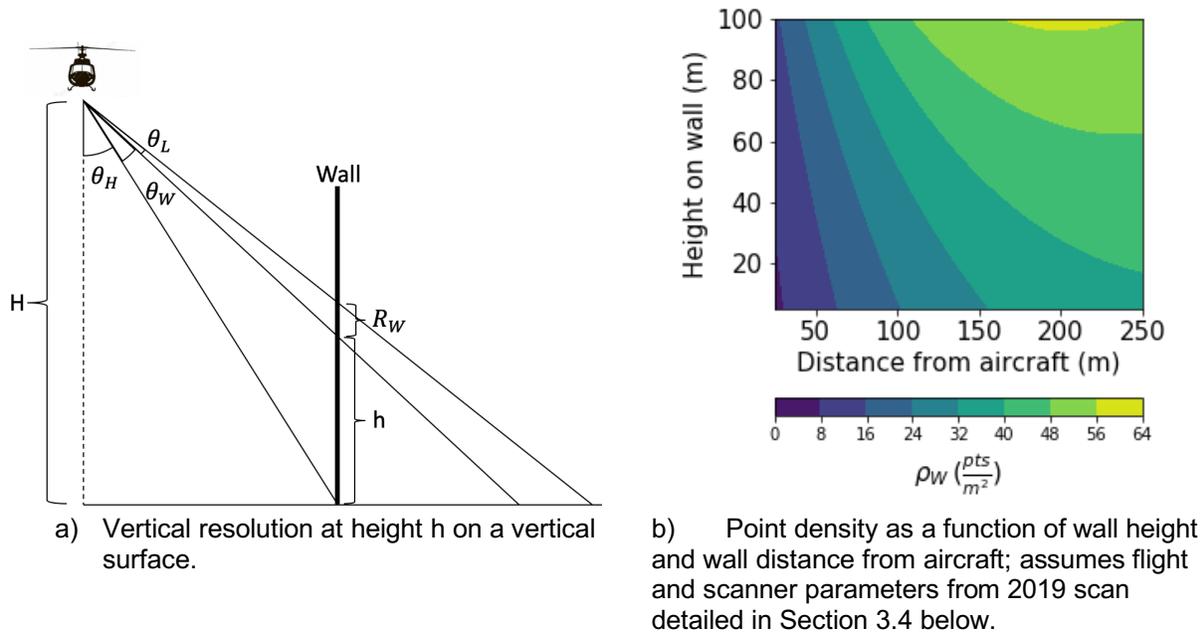

a) Vertical resolution at height h on a vertical surface.

b) Point density as a function of wall height and wall distance from aircraft; assumes flight and scanner parameters from 2019 scan detailed in Section 3.4 below.

Figure 5. Vertical resolution as a function of height on wall and wall distance from a helicopter.

### 3.1 Vertical point density

To fully capture the impact of the aircraft location on vertical density, this section extends the angle of capture analysis by Hinks et al. (2009) of section 2.3 to consider the vertical resolution at different wall heights. Figure 5a extends Figure 3 by introducing $R_W$, the vertical resolution on the vertical wall at height h. $R_W$ can be expressed in terms of $R_V$, $\theta_H$, and $\theta_W$:

$$R_W = R_V \frac{\sin^2(\theta_H)}{\sin^2(\theta_H + \theta_W)} \qquad \text{(eq 11)}$$

Eq (11) can also be written in terms of point density:

$$\rho_W = \rho_V \frac{\sin^2(\theta_H + \theta_W)}{\sin^2(\theta_H)} = \rho_N \frac{\tan(\theta_H + \theta_L)\sin^2(\theta_H + \theta_W)}{\tan^2(\theta_H)} \qquad \text{(eq 12)}$$

As stated in Section 2, point density is lowest at the base of the wall, which agrees with Eq (12): $\min(\rho_W) = \rho_V$ in the range $0° \leq \theta_H + \theta_W \leq 90°$, and $\rho_V$ corresponds to $\theta_W = 0°$. Importantly, the density gradient on the wall is larger for lower altitudes AGL, as both $\theta_H$ and $\theta_W$ increase. This implies that angle of capture effects are more significant for modern, high-density aerial LiDAR scans. Figure 5b shows the dependency of $\rho_W$ on wall height and offset distance.

Aggregate nominal point density for vertical surfaces ($ANPD_V$) is defined identically to its horizontal counterpart, $ANPD_H$, but applied to vertical surfaces. While $ANPD_H$ is commonly reported for aerial LiDAR datasets, $ANPD_V$ is not. $ANPD_V$ cannot be directly calculated from $ANPD_H$, as the two metrics are not directly proportional due to the impact of angle of capture, among other factors. To address this gap, this paper introduces $\eta_{HV}$, the ratio of horizontal to vertical density which is defined as:

$$\eta_{HV} = \frac{ANPD_H}{ANPD_V} \qquad \text{(eq 13)}$$

The term $\eta_{HV}$ is greater than one for nadir scanners operated in the $0° \leq \theta_H \leq 45°$ range, as depicted in figure 4. Notably, $ANPD_V$, and $\eta_{HV}$ do not explicitly consider the wall height, as they are aggregated over the entire vertical surface. The upper portions of the tallest building in the study area could be used to predict the theoretical maximum density. Thus, the overall vertical yield will be generally lower than the theoretical maximum due to surface roughness and the presence of windows, among other things. In contrast, horizontal planes have significantly fewer, highly reflective surfaces where returns fail to generate.

The equations in sections 2.3 and here assume a nadir scanner. Similar equations can be derived for other types of scanners, such as oblique scanners (where the scanner does not face directly downward from the aircraft) by adjusting $\theta_H$ and $\theta_L$ for the scanning angle and pattern. Unmanned aerial vehicles (UAV) used in scanning can fly below building height and utilize scan angles $\theta_H \geq 90°$ measured from nadir. The analysis demonstrated herein was devised as an extensible framework that can be adjusted for the rapidly growing variety of scanner types and platforms that are becoming increasingly available.

*3.2 Accuracy*

While knowing the overall accuracy of a point cloud is helpful for single pass aerial LiDAR, the impact of overlapping flight passes must be considered in high-density datasets. The value of increased point density obtained from multiple flight passes must be weighed against the potential of increased error due to misalignment of multiple flight passes.

The proposed metrics decompose local accuracy for point clouds composed of multiple, overlapping flight strips by an extension of the approach introduced in Latypov (2002). However, unlike Latypov (2002), who defined similarity metrics for pairs of overlapping, disparate point clouds, the accuracy metrics introduced herein decompose the error of a single point cloud into cross-pass and within-pass components. The focus here is error attribution. Additionally, the metrics introduced herein generalize to any surface orientation (i.e., vertical, horizontal, canted) and any number of flight passes, whereas Latypov (2002) only considered pairs of flight passes on horizontal surfaces. Figure 6 illustrates the proposed approach.

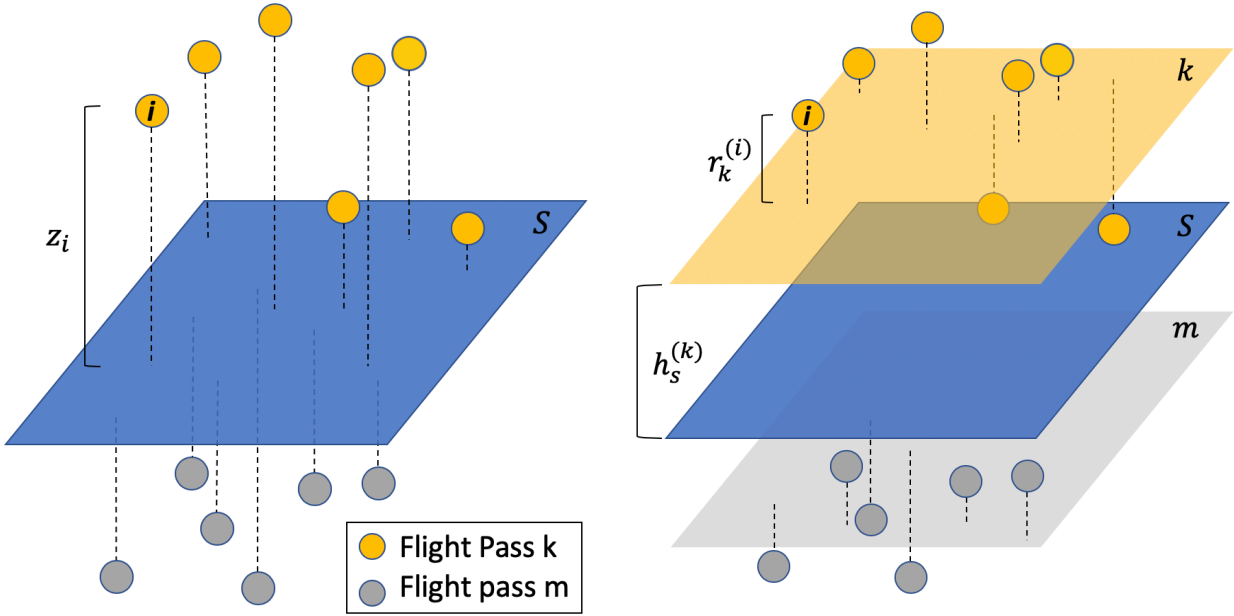

a) Plane S is fitted to the points collected from two independent flight passes, k and m; $z_i$ is the distance from each individual point to plane S

b) The term $r_k^{(i)}$ is the distance from plane k, which was fitted to only data from flight pass k, while $\widehat{h_S^{(k)}}$ is the distance between planes k and S

Figure 6. Decomposition of distance $z_i$ (left) into $\widehat{h_S^{(k)}}$ and $r_k^{(i)}$ (right) for point i from flight pass k. Two flight passes are shown but the approach is applicable to any number of passes.

Beginning with RMSE of a point cloud with respect to a flat surface, the distance $z_i$ of point i from the surface can be written as:

$$z_i = \widehat{h_S^{(k)}} + r_k^{(i)} \qquad \text{(eq 14)}$$

where, $\widehat{h_S^{(k)}}$ is the average distance of points in flight pass k from the surface S. Here, the known surface G is replaced with S, a plane fit to the points from all flight passes, and $\widehat{h_S^{(k)}}$ is measured in the direction orthogonal to S, rather than Latypov's vertical direction. This generalization allows for the application to non-horizontal surfaces. The term $r_k^{(i)}$ is the deviation of $z_i$ from $\widehat{h_S^{(k)}}$ for a point i in flight pass k.

Substituting Eq (14) into the equation for RMSE:

$$RMSE_S = \sqrt{\frac{\sum z_i^2}{N-1}} = \sqrt{\frac{\sum_k \sum_{i \in k}\left(h_S^{(k)} + r_k^{(i)}\right)^2}{N-1}} = \sqrt{\frac{\sum_k (n_k h_S^{(k)^2} + n_k MSE_k)}{N-1}} \qquad \text{(eq 15)}$$

Within each flight pass k, the sum of the residuals is zero, so the cross-term $\sum_k 2 h_S^{(k)} r_k^{(i)}$ is zero. Here, $RMSE_S$ is the square root of the weighted average over all flight passes of the average orthogonal offset, $\widehat{h_S^{(k)}}$, and the mean squared error of the points in the flight pass, $MSE_k$. Hence, RMSE can be decomposed via two new metrics, W and C:

$$W = \sqrt{\frac{1}{N-1} \sum_k n_k MSE_k}$$

$$C = \sqrt{\frac{1}{N-1} \sum_k n_k h_S^{(k)^2}}$$

$$RMSE_S = \sqrt{C^2 + W^2} \qquad \text{(eq 16)}$$

W is the weighted average MSE over all flight passes and represents the portion of error attributable to within-pass random error. The term W corresponds to the precision of surface measurements and is likely to be dependent on the quality of the scanning equipment (scanner, GPS, IMU) and the roughness of the selected "true surface". In contrast, C is the weighted average of $\widehat{h_S^{(k)}}$ over all flight passes and represents the error attributable to misalignment across flight passes. The parameter C is mostly dependent on the quality of system calibration, which often can be reduced (albeit not entirely eliminated) by strip adjustments after data collection. Notably, Eq (16) is an exact decomposition of RMSE: because $\widehat{h_S^{(k)}}$ and $r_k^{(i)}$ are defined orthogonally to the surface S, no approximation is introduced.

*3.3 Work flow*
The main components of the proposed workflow involve the surface selection and patch sampling (Figure 7a). The recommended process involves choosing planar, smooth, unobstructed, and opaque surfaces (Figure 7b). Once a surface is chosen, the area to be sampled was bounded precisely by selecting three points near the corners of the area, fitting a plane to those points, and defining a rectangular boundary lying in the plane, as shown in Figure 7b. Identifying and bounding the sample surfaces was performed manually, but the remainder of the workflow can be automated.

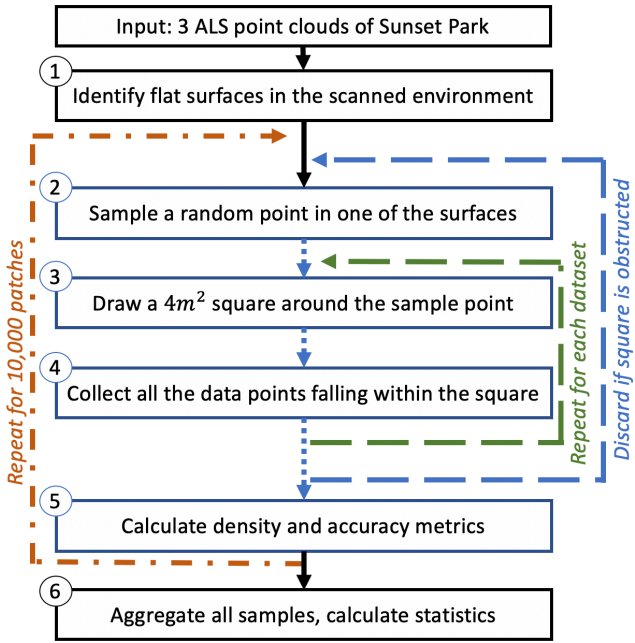
a) Surface selection and sampling processes.

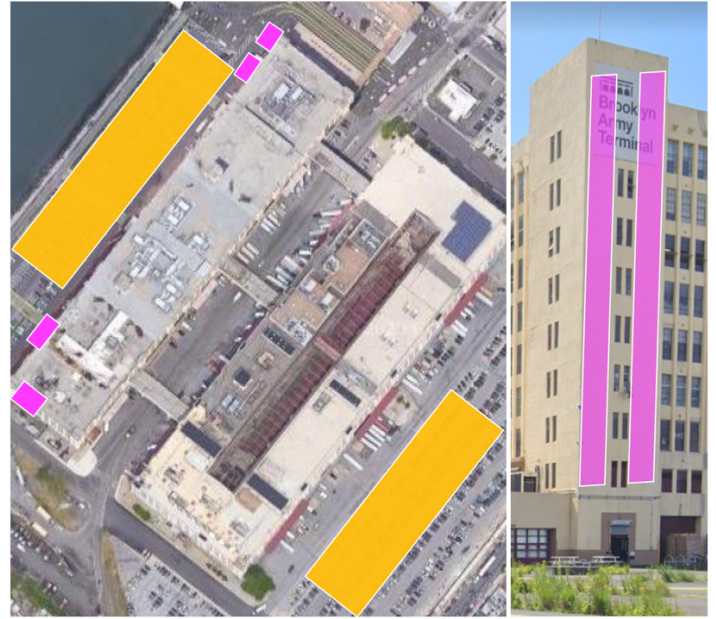
b) Sample surfaces around the Brooklyn Army Terminal in Sunset Park, Brooklyn, NY. Yellow rectangles indicate the horizontal surfaces. Pink rectangles indicate the vertical surfaces (Map, *Google Maps*).

Figure 7. Workflow and surface selection.

Points were sampled randomly from within the rectangular surface, and a square, sample patch was drawn around the point. Sample patch dimensions were devised to be (1) sufficiently large to ensure that patches include more than three points, (2) sufficiently small to fit the smallest dimension of the sample surface, and (3) the same for vertical and horizontal surfaces. Sample patches were discarded, if the patch contained unwanted obstructions (e.g., car, bench, wall ornamentation). Ten thousand patches were sampled for both horizontal and vertical surfaces. No restriction was placed to prevent patches from overlapping; while overlapping patches may introduce correlation across patches, the analyses performed herein do not rely on uncorrelated patches. The same patches were used for each dataset. Statistics were calculated over the random sampling.

3.4 Datasets

To demonstrate the usefulness of the proposed metrics, as well as the specific gains generated from targeted, high-density, municipal scans, a trio of recent scans (2014-2019) of a $1km^2$ area of Brooklyn, New York were compared. Buildings in the area are low- to medium-rise and do not exceed a height of $30m$. Each scan was conducted under different flight parameters and with different equipment (Figure 8). Each scan was commercially provided. Strip adjustments were performed on the 2019 scan. No post-processing information on 2014 and 2017 scans was available.

| | USGS 2014 (OCM Partners 2015) | NYC 2017 (NYC DOITT 2018) | Sunset Park 2019 (Laefer and Vo 2020) |
|---|---|---|---|
| Scan date | March/April 2014 | May 2017 | April 2019 |
| LiDAR scanner | Leica ALS70 | Leica ALS80 | Riegl LMS-Q680i |
| Pulse rate | 239kHz | 314kHz | 400kHz |
| FOV | 32° | 30° | 30° |
| Aircraft | Cessna 404/Cessna 310 (fixed wing) | Cessna 402C (fixed wing) | Bell 206 (rotary) |
| Flight speed | 278kph | 270kph | 93kph |
| Flight AGL | 2290m | 1800m | 300m |
| Swath sidelap | 30% | 60% | 77% |
| # of flight passes | 4 | 7 | 82 |
| Total flight time | 25 min | 2hr 6min | 5hr 0min |

a) Flight parameters for the three scans.

b) Flight altitude AGL relative to tallest building in scanned area.

Fig. 8. Flight parameters and AGL visualization for the three scans of Sunset Park, Brooklyn.

The 2014 scan was commissioned by the USGS and conducted by a third party contractor to evaluate storm damage and erosion of the local environment due to Hurricane Sandy. The 2017 scan was funded by New York City after being awarded a Disaster Recovery Community Development Block Grant related to Hurricane Sandy. These two scans covered large land areas and were conducted by fixed-wing aircraft at high speeds and large altitudes AGL. In contrast, the 2019 NYU-funded scan was conducted by a helicopter at a lower speed and a lower altitude AGL. Figure 9 provides a selective visualization of objects in these scans and demonstrates the difficulty of identifying sub-building scale objects in low density scans.

For the datasets tested herein, portions of a largely empty parking lot were used to sample the horizontal data and flat, unadorned portions of building walls with no windows were used as vertical surfaces (Fig. 7b). The goal of this paper was to document the built environment, so surfaces were chosen to be free of vegetation. The horizontal data came from $4,460 m^2$ of asphalt from 2 parking lots. The vertical data were taken from $350 m^2$ from 4 vertical strips along a single building. Horizontal samples were discarded if a car was present, resulting in the resampling of roughly 30% of the initial points. Wall surfaces were unadorned and free of windows, so no vertical points were discarded.

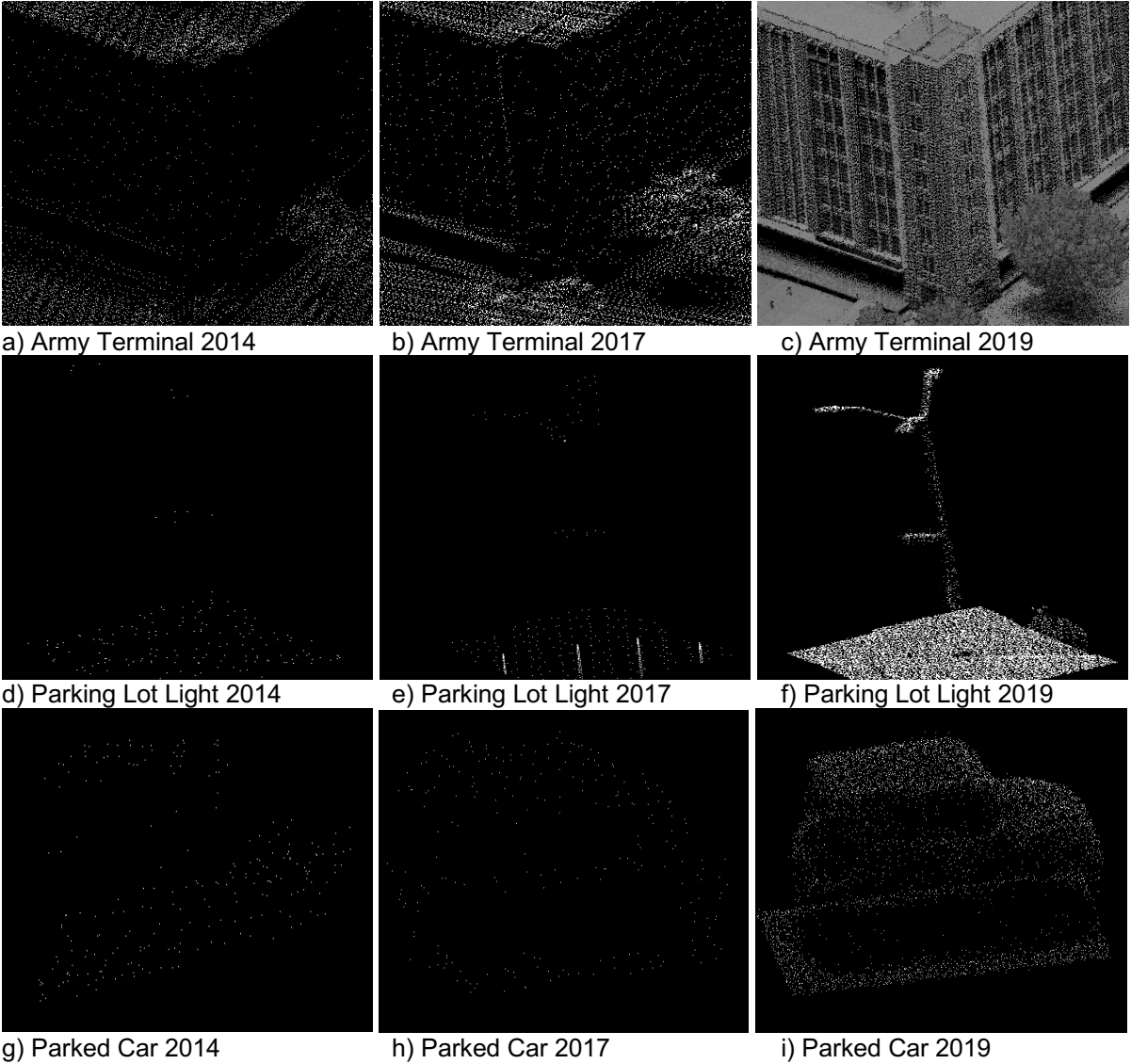

Figure 9. Similar objects shown for each of the three scans. Points are shaded by the intensity captured. The vehicles are not the same from scan to scan, but similar models were selected from the same area of each scan.

## 4. RESULTS
The suite of five proposed metrics were applied to the three datasets described in section 3.4. Point density and accuracy on the horizontal and vertical surfaces are reported for the three scans. Vertical point density for the 2019 scan are compared to the theoretical predictions at different heights along the tallest wall in the study area.

*4.1 Point density generally*
The horizontal densities of the 2014 and 2017 scans (7 $pts/m^2$ and 11 $pts/m^2$, respectively) are similar to each other, but the 2019 scan at 510 $pts/m^2$ is more than an order of magnitude denser. This was intentionally achieved through use of improved sensor hardware, lower flight altitude AGL, slower flight velocity, and more overlapping flight passes (6-7 times more).

While the horizontal density increases progressively for each subsequent scan, vertical density increases much faster, which is evident from the declining horizontal to vertical density ratio, $\eta_{HV}$: 46.1 in 2014, 16.5 in 2017, and 10.3 in 2019. The improvement in $\eta_{HV}$ in newer scans is directly attributable to the specifics of the flight plan; specifically, the lower altitudes AGL of the later flights, as shown by the theoretical values (Table 2). Holding all else constant, a given wall height causes a larger $\theta_W$ for lower altitudes AGL. Larger $\theta_W$ increases the $\frac{\rho_W}{\rho_N}$ ratio [see Eq (12)] and reduces $\eta_{HV}$.

Average densities are lower than theoretical densities because sample surfaces may only be partially visible to some flight passes. This is particularly true of vertical surfaces, where a flight path may only generate a few points on the surface because the angle between path and the surface normal is highly acute. Average point density for parallel flight paths is provided in Table 2 and is a more direct comparison to the theoretical density, as Eq (12) assumes a parallel flight path.

|  | USGS 2014 (OCM Partners 2015) | NYC 2017 (NYC DOITT 2018) | Sunset Park 2019 (Laefer and Vo 2020) |
|---|---|---|---|
| *Horizontal* | | | |
| Theoretical point density per flight pass ($pts/m^2$) | 2.58 | 4.46 | 33.27 |
| Av. point density per flight pass | 2.26 | 3.64 | 25.78 |
| Actual aggregate point density ($pts/m^2$) | 6.50 (SD: 2.19) | 10.93 (SD: 2.45) | 510.49 (SD: 24.80) |
| Av. number of overlapping flight passes | 2.88 | 3.00 | 19.80 |
| *Vertical* | | | |
| Theoretical point density per flight pass ($pts/m^2$) | 1.15 | 1.98 | 14.83 |
| Av. point density per flight pass | 0.24 | 1.02 | 5.24 |
| Av. point density - parallel flight passes | - | 1.81 | 14.48 |
| Actual aggregate point density ($pts/m^2$) | 0.14 (SD: 0.12) | 0.66 (SD: 1.39) | 49.74 (SD: 10.24) |
| Av. number of overlapping flight passes | 0.60 | 0.65 | 9.49 |
| $\eta_{HV}$ | 46.08 | 16.49 | 10.26 |

Table 2. Density metrics for Brooklyn, NY LiDAR Scans. Horizontal theoretical point density is calculated at nadir. Vertical theoretical point density is calculated as the average density over wall heights from 0m to 30m for a wall at $\theta_H = 27.5°$ (max angle for which top of wall is fully within operating range).

*4.2 Point density by wall height*
Figure 10 shows the actual vertical data for each of the 3 datasets. The density by height for each is quantified in Fig. 11. This compares the 2017 and 2019 results to the theoretical point density predictions of Eq (12); the metadata for the 2014 lacked the explicit trajectory information needed for this calculation. The 2017 and 2019 flight paths selected ran parallel to the wall face at a constant offset, with the entire wall face in the scanner's operating range.

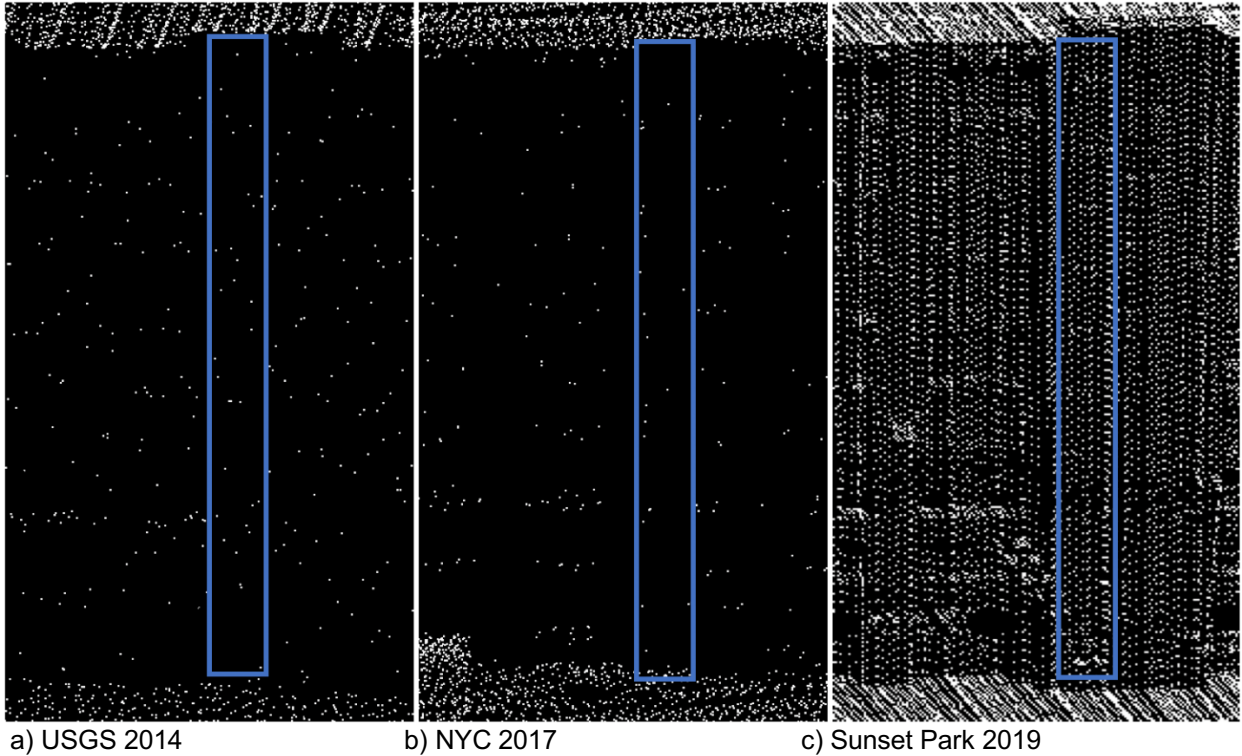

a) USGS 2014　　　　　　　b) NYC 2017　　　　　　　c) Sunset Park 2019

Figure 10. Point cloud of vertical wall of Army Terminal Building in each scan. Blue rectangle indicates a sample surface.

The 2019 results largely agree with the predictions, with a mean absolute error (MAE) of only $0.53\ pts/m^2$ and mean absolute percentage error (MAPE) of 3.6%. In contrast, the 2017 density is quite erratic due to random noise in the very small number of points. This is evident from the MAE of $0.10\ pts/m^2$ and MAPE of 49.2%. Eq (12) also correctly predicted a much larger impact from the angle of capture for the 2019 scan than the 2017 scan due to the lower flight altitude AGL and, hence, larger angles $\theta_H$ and $\theta_W$.

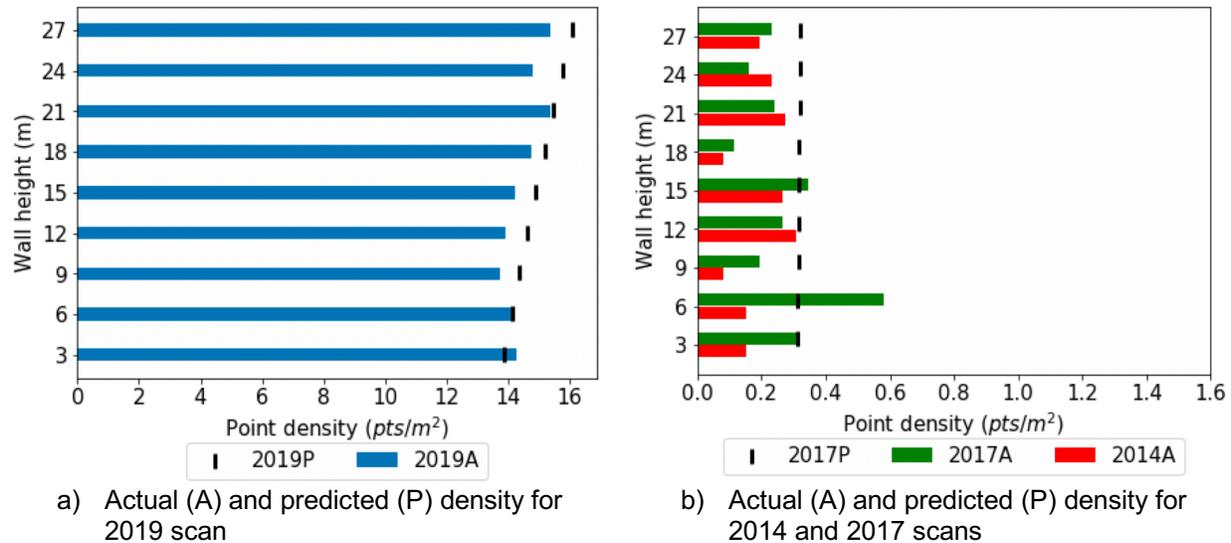

a) Actual (A) and predicted (P) density for 2019 scan

b) Actual (A) and predicted (P) density for 2014 and 2017 scans

Fig. 11. Comparison of actual versus predicted vertical densities at different wall heights for a single flight path in each scan. Ground features obstruct density estimates below $3m$. The predicted densities are calculated from Eq (12) and flight parameters. 2019 path flown at $300m$ AGL and $97m$ horizontal offset from wall; 2017 path flown at $1800m$ AGL and $315m$ horizontal offset from wall. Flight path not available for 2014 scan, so density could not be predicted.

*4.3 Accuracy*

Table 3 reports the accuracy metrics for horizontal and vertical surfaces for the three scans. Horizontal accuracy consistently improved in the more recent scans. The error decomposition in Table 4 shows that this improvement is due to both lower cross-pass error ($C_H$) and lower within-pass error ($W_H$). For all three scans, the C/W ratio of 0.42-0.52 indicates that the majority of error for horizontal surfaces was attributable to within-pass error. Critically, $C_H$ decreases in more recent scans despite the inclusion of significantly more scan in the 2019 [i.e. 19.8 flight passes (on average) versus only 3.0 in 2017 and 2.9 in 2014]. The decrease in cross-pass ratio indicates that the higher number of passes in 2019 is less impactful on cross-pass error than other factors (e.g., system calibration, time between flights). The within-pass error steadily decreases over the years from a high of $0.019m$ in 2014 to $0.004m$ in 2019. Notably, cross-pass error remains in the 2019 scan despite post-flight strip adjustments.

Vertical RMSE is 218% higher than horizontal RMSE for the 2019 scan and nearly as high as the horizontal RMSE for the 2014 scan. Interestingly, the vertical C/W ratio of 1.68 indicates that, unlike the findings for horizontal surfaces, the majority of error for vertical surfaces is attributable to cross-pass error. Specifically, $C_V$ is 539% higher than $C_H$ for the 2019 scan (with fewer than half as many flight passes), while the average single flight pass RMSE is only 112% higher for vertical surfaces.

|  | USGS 2014 (OCM Partners 2015) | NYC 2017 (NYC DOITT 2018) | Sunset Park 2019 (Laefer and Vo 2020) |
|---|---|---|---|
| *Horizontal* | | | |
| Av. number of overlapping flight passes | 2.880 | 3.000 | 19.800 |
| $RMSE_H$ (m) | 0.043 | 0.017 | 0.010 |
| $C_H$ (m) | 0.019 | 0.007 | 0.004 |
| $W_H$ (m) | 0.038 | 0.015 | 0.009 |
| C/W ratio | 0.512 | 0.456 | 0.427 |
| *Vertical* | | | |
| Av. number of overlapping flight passes | - | - | 9.490 |
| $RMSE_V$ (m) | - | - | 0.032 |
| $C_V$ (m) | - | - | 0.024 |
| $W_V$ (m) | - | - | 0.019 |
| C/W ratio | - | - | 1.286 |

Table 3. Local accuracy comparison of LiDAR scans for Brooklyn, NY. The 2014 and the 2017 scans had vertical point densities below 1.5 $pts/m^2$ (fewer than 6 points per $4m^2$ sample square) and rarely more than one flight pass per sample square. For this reason, both were excluded from the vertical accuracy analysis.

## 5. DISCUSSION

Two additional points are relevant to the discussion of new metrics for high density aerial LiDAR scanning: metric sensitivity to sample patch size and robustness in its applicability to other locations.

|  | $1m^2$ | $2m^2$ | $4m^2$ |
|---|---|---|---|
| **DENSITY** | | | |
| *Horizontal* | | | |
| Point density ($pts/m^2$) | 513.23 (SD: 24.66) | 512.59 (SD: 23.20) | 510.49 (SD: 24.80) |
| Av. point density per flight pass ($pts/m^2$) | 25.92 | 25.89 | 25.78 |
| *Vertical* | | | |
| Point density ($pts/m^2$) | 49.73 (SD: 11.45) | 49.52 (SD: 9.88) | 49.74 (SD: 10.24) |
| Av. point density per flight pass ($pts/m^2$) | 6.16 | 5.69 | 5.24 |
| $\eta_{HV}$ | 10.32 | 10.35 | 10.26 |
| **ACCURACY** | | | |
| *Horizontal* | | | |
| $RMSE_H$ (m) | 0.010 | 0.0096 | 0.0098 |
| $C_H$ (m) | 0.005 | 0.0042 | 0.0038 |
| $W_H$ (m) | 0.008 | 0.0085 | 0.0089 |
| *Vertical* | | | |
| $RMSE_V$ (m) | 0.028 | 0.029 | 0.029 |
| $C_V$ (m) | 0.024 | 0.025 | 0.025 |
| $W_V$ (m) | 0.014 | 0.015 | 0.015 |

Table 4. Density and accuracy metrics for the 2019 scan using different sample patch sizes.

*5.1 Patch size sensitivity analysis*

The results reported in Section 4 were calculated using a $4m^2$ patch size. Patch selection size was restricted by the availability of vertical surfaces uninterrupted by windows. To check if patch size selection impacted the results, the analysis was rerun on the 2019 data considering patches of $1m^2$ and $2m^2$. Patch size was found to have no identifiable impact on density and accuracy metrics across these scales (Table 4).

*5.2 Robustness: Application to 2015 Dublin scan*

To test the applicability of the proposed metrics to other data sets, they were applied to a 2015 high density scan of Dublin, Ireland (Laefer et al., 2017). That scan utilized the same LiDAR scanner (Riegl Q680i) as the Sunset Park 2019 scan, and combined 44 flight paths flown at a height of $300m$ AGL and a speed of 68 $kph$, but only at 67% overlap and achieved an aggregate point density of 348 $pts/m^2$ over $1.5km^2$ of central Dublin. Strip adjustments were made to both datasets.

|  | **Dublin City 2015 (Laefer et al. 2017)** | **Sunset Park 2019 (Laefer and Vo 2020)** |
|---|---|---|
| **DENSITY** |  |  |
| *Horizontal* |  |  |
| Point density ($pts/m^2$) | 290.96 (SD: 50.63) | 510.49 (SD: 24.80) |
| Av. number of overlapping flight passes | 8.40 | 19.80 |
| Av. point density per flight pass ($pts/m^2$) | 34.64 | 25.78 |
| *Vertical* |  |  |
| Point density ($pts/m^2$) | 24.84 (SD: 5.99) | 49.74 (SD: 10.24) |
| Av. number of overlapping flight passes | 4.17 | 9.49 |
| Av. point density per flight pass ($pts/m^2$) | 5.96 | 5.24 |
| $\eta_{HV}$ | 11.71 | 10.26 |
| **ACCURACY** |  |  |
| *Horizontal* |  |  |
| Av. number of overlapping flight passes | 8.40 | 19.80 |
| $RMSE_H$ (m) | 0.013 | 0.010 |
| $C_H$ (m) | 0.009 | 0.004 |
| $W_H$ (m) | 0.009 | 0.009 |
| C/W ratio | 1.002 | 0.427 |
| *Vertical* |  |  |
| Av. number of overlapping flight passes | 4.17 | 9.49 |
| $RMSE_V$ (m) | 0.022 | 0.029 |
| $C_V$ (m) | 0.012 | 0.025 |
| $W_V$ (m) | 0.017 | 0.015 |
| C/W ratio | 0.723 | 1.678 |

Table 5. Density and accuracy metrics for the 2015 Dublin scan and 2019 Sunset Park scan.

Five sample surfaces were used: (1) a parking lot behind Leinster House (horizontal), (2) Barnardo Square next to City Hall (horizontal), (3) a paved walking area in St. Stephen's Green (horizontal), and (4) the south and (5) west walls of La Touche House (vertical). Notably, the surface materials are not necessarily identical to those of the Sunset Park scans with respect to roughness and reflectivity.

Table 5 provides the density and accuracy metrics for the Dublin scan. The similar horizontal-to-vertical density ratios (11.71 for Dublin and 10.26 for Brooklyn) are reflective of the two scans having been performed under similar flight parameters (altitude AGL and velocity) and with the same LiDAR scanner. The lower density in the Dublin scan for both horizontal (43%) and vertical (50%) surfaces reflects the significantly fewer flight passes compared to the Sunset Park scan (41 versus 82).

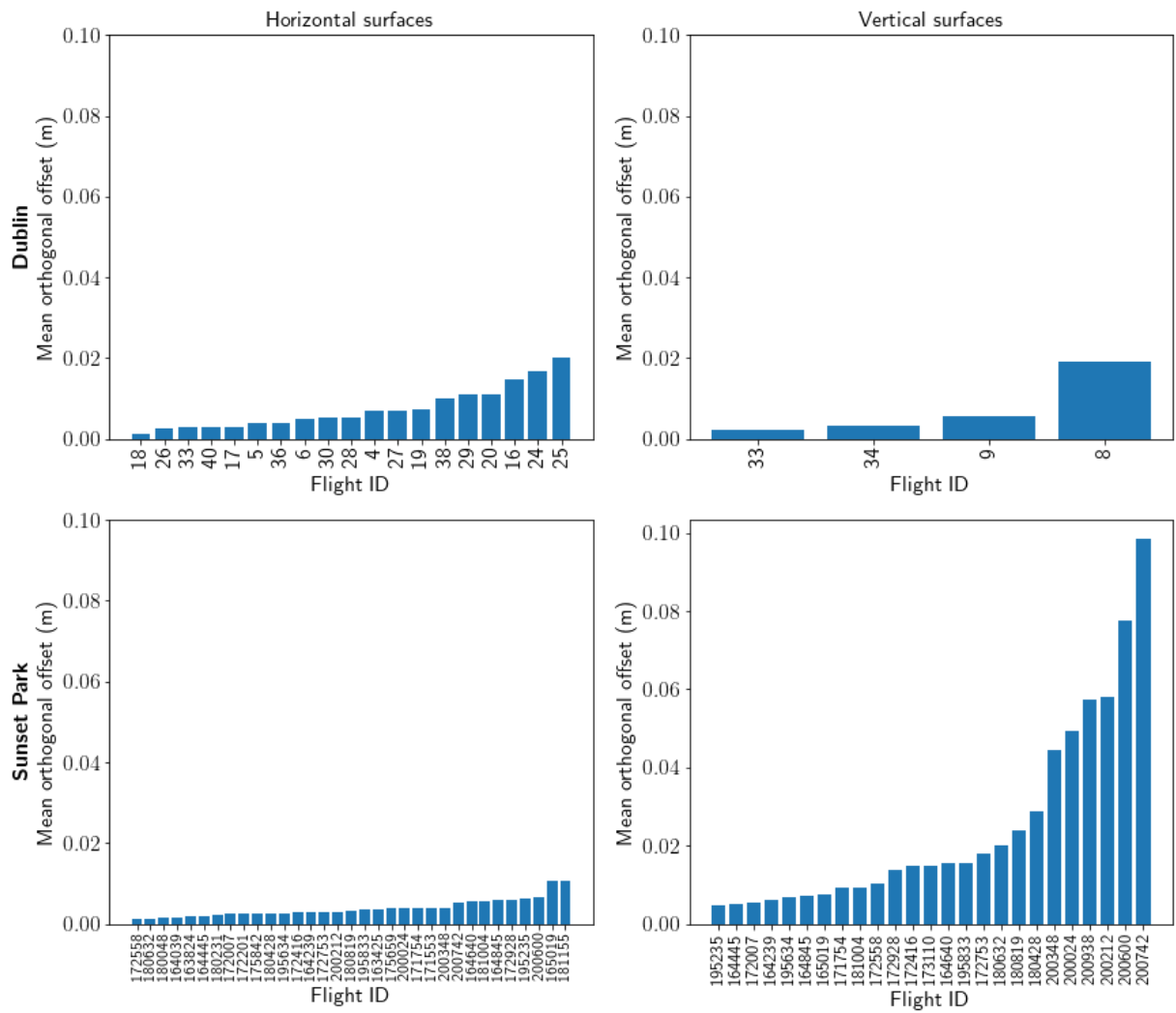

Figure 12. Mean orthogonal offset relative to fitted surface S for each flight pass over the samples of both datasets and both dimensions. Two outlier flight passes are omitted from Dublin vertical, because they each had only two points on the surface.

The datasets exhibited very similar within-pass error on both horizontal surfaces ($0.009m$) and vertical surfaces ($0.017m \ and \ 0.015m$ for the Dublin and Sunset Park scans, respectively). Similar W values, likely due to comparable equipment and flight parameters, indicate robustness of the W metric to different datasets and surface materials. In contrast, the cross-pass error levels differed more across the two datasets: $C_V$ was twice as high for Sunset Park ($0.025m$ versus $0.012m$). However, $C_H$ is higher for Dublin ($0.009m$ versus $0.004m$). Figure 12 delves into the causes of this discrepancy by plotting the mean orthogonal offset, or:

$$\frac{1}{K}\Sigma_k \, abs(\widehat{h_S^{(k)}}) \qquad \text{(eq 17)}$$

for k flight passes. Mean absolute height is an indication of how far a single flight pass tends to differ from the other flight passes with similar coverage. For horizontal surfaces, the distribution of mean heights across flight paths is similar for the two scans, and the larger $C_H$ value for the Dublin scan is simply due to consistently larger mean heights (Fig. 12). For vertical surfaces, six flight passes for Sunset Park cause the majority of $C_V$. Those six flight passes are all in the same direction (northeast-southwest), and create very acute angles with the normal of the sample surfaces. These flight passes are depicted by the dotted black lines in Figure 13. None of the flight passes in the Dublin scan are as acute, which could be the reason that none of them have extreme orthogonal offsets, unlike that which occurs in the Sunset Park data.

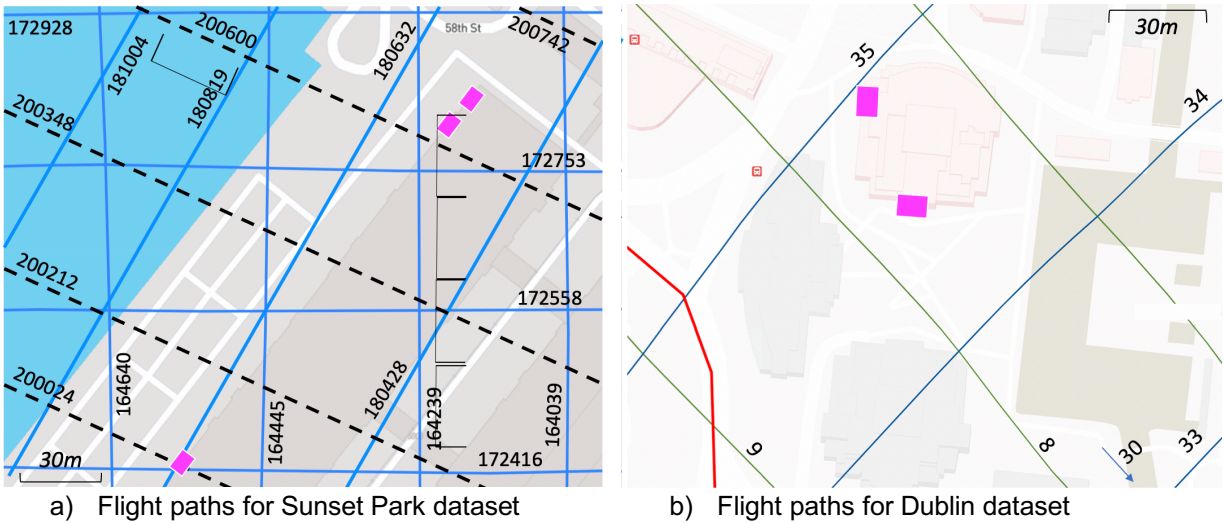

a) Flight paths for Sunset Park dataset      b) Flight paths for Dublin dataset

Figure 13. Flight paths indicated by overlaid lines and vertical sample surfaces indicated with pink boxes. Dotted black lines in Sunset Park figure indicate flight passes at extreme acute angles relative to the sample surface normals.

Figure 14 compares actual and predicted point density at various wall heights for a single flight pass from both the Dublin and Sunset Park scans. The predicted densities are calculated from Eq (12) and the flight parameters. The horizontal offset was $97m$ for the Sunset Park flight pass and $110m$ for the Dublin flight pass, and both passes were flown at an altitude of $300m$ AGL. Point density increased at higher points on the wall for both datasets, as predicted. The Dublin scan, in fact, fits the prediction more closely than the Brooklyn scan, with a MAE of $0.39 \, pts/m^2$ and MAPE of 3.8%, though the actual densities were systematically lower than the predictions. This discrepancy is likely attributable to local wind conditions.

The newly introduced metrics enabled a richer comparison of two, modern, high-density urban scans flown for the same mission aims. The similarities of flight parameters and equipment are apparent from the similar $\eta_{HV}$ ratios and within-pass error, W. The cause of the difference in vertical accuracy was apparent from the error decomposition: acute angles between flight paths and vertical surface normals in the Sunset Park dataset increased the cross-pass error.

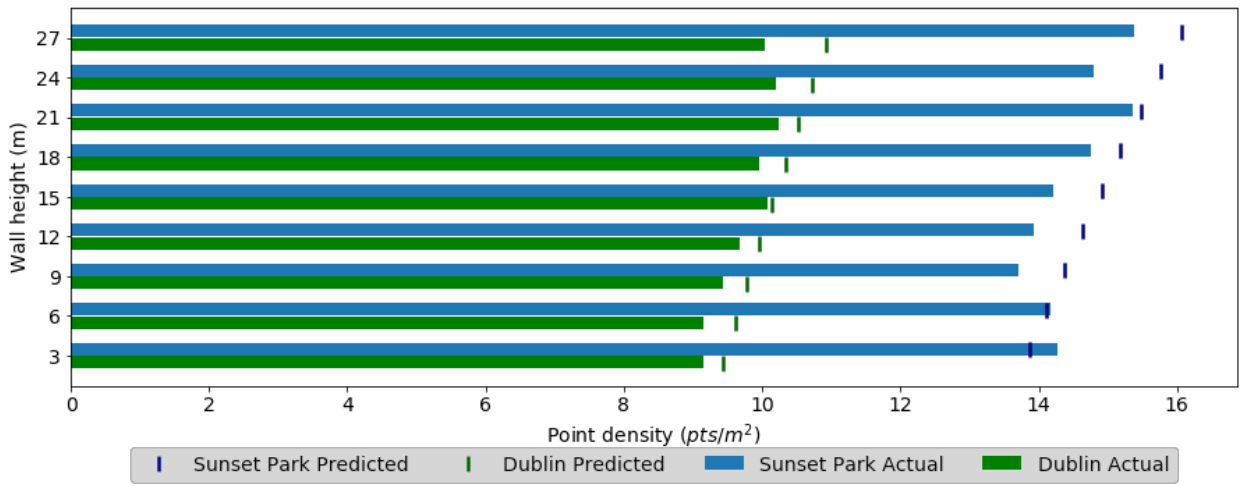

Fig 14 Comparison of predicted (vertical lines) and actual (bars) vertical densities at different wall heights for a single flight path in the scans of Dublin and Sunset Park (2019).

## 6. CONCLUSIONS

This paper introduces five new metrics for high-density aerial LiDAR datasets (for assessing existing datasets and future flight planning). These metrics describe density and accuracy for vertical surfaces and the impact of combining numerous, overlapping flight passes on accuracy. The increased desire for better vertical data capture in urban environments and the reliance on multi-pass missions for more complete coverage motivate the need for these metrics. This paper also quantifies the impact of angle of capture on vertical density. Accounting for this impact should enable better density prediction during flight planning.

Beyond the introduced metrics, this paper makes three significant contributions. The first is the introduction of a vertical density estimation equation that considers both sensor position and target wall height. By explicitly including the angle of capture, the vertical density (which cannot be directly estimated from the horizontal density) can now be reliably predicted for specific buildings of interest. This enables a much more tailored flight mission if a specific minimum density is required for visualization or other downstream applications. The second contribution is the use of cross-pass error as a meaningful parameter to compare high-density, urban point clouds. For point clouds generated under similar flight parameters, the cross-pass error will fully identify any differences in accuracy contributable to the execution of the flyover. Lastly, the paper identifies and quantifies the disproportionate impact of modern, optimized flight missions on vertical density (positive) and accuracy (negative). The detrimental impact on accuracy can be reduced by avoiding a reliance on highly acute flight paths for documenting buildings of particular interest. Understanding these factors can improve future mission planning and subsequent data processing. The verification and applicability of the proposed metrics were demonstrated on four recent, aerial LiDAR scans at two urban locations.

Several of the concepts presented in this paper warrant future consideration. Foremost is the need to test these metrics on datasets representing a broader range of flight parameters, equipment calibration, and surface compositions than what was compared herein. In order to predict vertical densities more generally, the angle of capture analysis should be extended to off-nadir scanners and various scan patterns.

ACKNOWLEDGEMENTS
The authors would like to thank Dr. Anh-Vu Vo for his input on the Sunset Park dataset and Tuck Mapping for conducting and collaborating on the 2019 Sunset Park scan.

FUNDING
This work was funded by the National Science Foundation (award 1940145) and the Data Science and Software Services program at New York University, which is funded through the Gordon & Betty Moore Foundation (#3836) and the Alfred P. Sloan Foundation (G-2016-7191).